\documentclass[conference]{IEEEtran}

\IEEEoverridecommandlockouts
\def\BibTeX{{\rm B\kern-.05em{\sc i\kern-.025em b}\kern-.08em
		T\kern-.1667em\lower.7ex\hbox{E}\kern-.125emX}}

\usepackage{dblfloatfix}    
\usepackage{cite}
\usepackage{amsmath,amssymb,amsfonts}
\usepackage{bbm}

\usepackage{graphicx}
\usepackage{textcomp}
\usepackage{xcolor}
\usepackage{colortbl} 

\usepackage{url} 

\usepackage{bm}
\usepackage[normalem]{ulem} 
\usepackage{supertabular}
\usepackage{amsthm}
\theoremstyle{definition}

\usepackage[linesnumbered,ruled,vlined]{algorithm2e}
\usepackage{algorithmicx} 
\usepackage{pbox}
\usepackage{float} 
\usepackage{subcaption} 
\usepackage{epstopdf}

\usepackage{boldline,multirow}
\usepackage{array}

\usepackage{pbox}

\usepackage{dblfloatfix}

\begin{document}
\setlength{\abovedisplayskip}{2pt}
\setlength{\belowdisplayskip}{2pt}


\title{Decentralized Uplink Adaptive Compression for Cell-Free MIMO with Limited Fronthaul
	\thanks{The authors would like to acknowledge the support of Ericsson Canada and the Natural Sciences and Engineering Research Council of Canada.}
}

\author{\IEEEauthorblockN{Zehua~Li, Jingjie~Wei and Raviraj~Adve}
	\IEEEauthorblockA{{Dept. of Electrical and Computer Engineering} \\
		{University of Toronto}\\
		Toronto, Canada \\
		\{samzehuali.li, peter.wei\}@mail.utoronto.ca, rsadve@ece.utoronto.ca}
}


%
%

\maketitle 


\begin{abstract}
We study the problem of uplink compression for cell-free multi-input multi-output networks with limited fronthaul capacity. In compress-forward mode, remote radio heads (RRHs) compress the received signal and forward it to a central unit for joint processing. While previous work has focused on a transform-based approach, which optimizes the transform matrix that reduces signals of high dimension to a static pre-determined lower dimension, we propose a rate-based approach that simultaneously finds both dimension and compression adaptively. Our approach accommodates for changes to network traffic and fronthaul limits. Using mutual information as the objective, we obtain the theoretical network capacity for adaptive compression and decouple the expression to enable decentralization. Furthermore, using channel statistics and user traffic density, we show different approaches to compute an efficient representation of side information that summarizes global channel state information and is shared with RRHs to assist compression. While keeping the information exchange overhead low, our decentralized implementation of adaptive compression shows competitive overall network performance compared to a centralized approach.

\end{abstract}

%

%
%
%
\section{Introduction} \label{section:intro}
With tremendous growth in the wireless connectivity market, service providers are constantly seeking ways to deliver higher data rates to denser populations. One key technology envisioned is cell-free multi-input multi-output (MIMO)~\cite{Björnson_2020} networks. In the uplink of a cell-free MIMO system, users are jointly served by multiple remote radio heads (RRHs) that forward their received signals to a central processing unit (CPU) for joint processing, which is often referred as compress-and-forward~\cite{Quek_Peng_Simeone_Yu_2017}. Two key benefits are that the RRH are close to users and that joint processing effectively addresses the issue of interference. Though this architecture eliminates the cell-edge users that are found in traditional cellular networks, it also introduces new problems. 

While many works have ignored the fronthaul, in practice, fronthaul capacity is limited and each RRH must compress its received signals, of dimension equal to the number of its antennas, before forwarding to the CPU. The associated quantization distortions may significantly degrade the data rate. A common strategy is transform-compress-forward~\cite{Liu_Yu_Simeone_2017}, first applying a dimension reduction transform matrix, and then compressing the lower-dimensional signal with a uniform quantizer. The focus has been designing the best transform matrix based on different levels of channel state information (CSI) required. For example, the authors in~\cite{wiffen_2021} use coordinate descent on conditional Karhunen-Loeve transform matrices if global CSI is available and an eigenvalue decomposition (EVD)-based matrix when only local CSI is available. While the former outperforms the latter, the associated CSI exchange overhead makes global methods impractical. As a result, the authors in~\cite{Sohrabi_2022,qiao_2023} proposed learning-based methods that aim to reach performance with global CSI while keeping the CSI sharing cost low. These transform-based methods tend to a priori pick the reduced dimension for all RRHs. However, this number has to be carefully selected based on he network traffic and fronthaul capacity, which may vary across RRHs.

The limited flexibility in the available literature motivates our formulation of a scheme that finds the compression and dimension simultaneously based on network conditions. Specifically, we propose a new \textit{rate-based} method, based on adaptive compression~\cite{vu_2017}, that aims to optimally split the fronthaul rate limit to each channel for every RRH by employing a rate allocation block (RAB). This not only finds a good compression rate allocation strategy but also, indirectly, determines how many dimensions to keep. We use mutual information based objectives to study the theoretical upper bound of what can be achieved with adaptive compression. 

The contributions of this paper are:
\begin{itemize}
	\item We formulate and solve the uplink compression rate allocation problem for both local and global information rate objectives.
	\item We propose decentralized adaptive compression with a RAB at every RRH. Using conditional mutual information, we decouple the global objective and formulate the generalized decentralized objective that for local optimization at each RRH. 
	\item To enable effective solutions in practice, we use two methods that utilize channel statistics and user traffic distribution to significantly decrease communication overheads of CSI sharing while maintaining comparable performance.
\end{itemize}

\section{System Model}\label{section:Model}
\subsection{Network Model} \label{subsection:network_model}
We consider the uplink of a distributed MIMO network operating in time-division duplex (TDD) mode. Our model comprises one CPU controlling a set of RRHs denoted by $\mathcal{R}$ with each RRH equipped with $M$ antennas. The set of users being served is denoted by $\mathcal{U}$ and each user is equipped with a single antenna. We expect abundant spatial resources where $|\mathcal{U} | < M | \mathcal{R} |$ so that all users are served simultaneously. The links between users and RRHs are called access links and the links between RRHs and CPU are called fronthaul links.

We use indices $r$ and $u$ to refer to RRHs and users respectively. The fronthaul for RRH $r$ has a limited capacity of $L_r$ bits/s/Hz. The uplink channel between user $u$ and RRH $r$ is modeled as $\mathbf{h}_{ru}  = \sqrt{\psi_{ru} \beta(d_{ru})} \mathbf{g}_{ru}$, where $\mathbf{g}_{ru} \in \mathbb{C}^{M}$ accounts for small-scale fading, modeled as unit-variance and Rayleigh; $\psi_{ru}$ and $\beta(d_{ru})$ denote the large-scale fading and pathloss over $d_{ru}$, the distance between RRH $r$ and user $u$.

\subsection{Signal Model}
Our transmission scheme is often referred to as compress-forward wherein the RRHs compress the received signals and then forward a quantized version to the CPU for processing. Ideally, this scheme reaps full cooperation gain through joint processing of signals from all RRHs. However, its performance is limited because the CPU processes the quantized signals and the level of distortion depends on the fronthaul capacity.

For compactness, we use $\mathbf{x} \in \mathbb{C}^{| \mathcal{U} |}$ to denote the transmissions from all users and $\mathbf{H}_r = [ \mathbf{h}_{ru_1} \mathbf{h}_{ru_2} \ldots \mathbf{h}_{ru_{|\mathcal{U}|}} ] \in \mathbb{C}^{M \times | \mathcal{U} |} $ as the channel matrix from RRH $r$ to all users. The vector $\mathbf{y}_r \in \mathbb{C}^{M}$ denotes the received signal at RRH $r$ while $\mathbf{z}_r \in \mathbb{C}^{M}$ denotes the compressed signal forwarded to the CPU by RRH $r$. The compress-forward signal model is given by
\begin{eqnarray} 
    \mathbf{y}_r & = &\mathbf{H}_{r} \mathbf{x} + \mathbf{n}_r \\
    \mathbf{z}_r & = & \mathbf{H}_{r} \mathbf{x} + \mathbf{n}_r + \mathbf{q}_r
\end{eqnarray}
where $\mathbf{n}_r \in \mathbb{C}^{M}$ denotes the white Gaussian noise and $\mathbf{q}_r \in \mathbb{C}^{M}$ the quantization distortion caused by compression. 

All users transmit at equal power with $\mathbf{Cov} ( \mathbf{x}  )=p\mathbf{I}_{|\mathcal{U}|}$ where $p$ is the power and $\mathbf{I}_{|\mathcal{U}|}$ denotes identity matrix of size $|\mathcal{U}|$. In addition to the noise covariance matrix $\mathbf{N}_r=\sigma_r^2 \mathbf{I}_{M}$, we define the quantization covariance matrix as $\mathbf{Cov} ( \mathbf{q}_r  ) = \mathbf{Q}_r=\mathrm{diag} \{ q_{r,m} \}_{m=1}^M$. Here, $q_{r,m}$ is a power term unlike $\mathbf{q}_r$.

We denote $\mathbf{z} = [ \mathbf{z}_{r_1}^T \mathbf{z}_{r_2}^T \ldots \mathbf{z}_{r_{|\mathcal{R}|}}^T ]^T \in \mathbb{C}^{M | \mathcal{R} |}$ as the concatenation of compressed signals received from all RRHs. We define the channel matrix for all channels in the network $\mathbf{H} = [ \mathbf{H}_{r_1}^T \mathbf{H}_{r_2}^T \ldots \mathbf{H}_{r_{|\mathcal{R}|}}^T ]^T \in \mathbb{C}^{M | \mathcal{R} | \times | \mathcal{U} | }$ in a similar manner. Hence, the signal received at the CPU can be written as 
\begin{equation} 
    \mathbf{z} = \mathbf{H} \mathbf{x} + \mathbf{n} + \mathbf{q}
\end{equation}
where $\mathbf{n}$ and $\mathbf{q}$ are defined accordingly with vertical concatenations and their covariance matrices $\mathbf{N}$ and $\mathbf{Q}$ are block diagonal concatenations of $\mathbf{N}_r$ and $\mathbf{Q}_r$ respectively.

\subsection{Capacity Formulation}
Using the signal model, we first list the mutual information expressions with respect to one RRH as 
\begin{equation} \label{eqn:compression_rate}
    I(\mathbf{y}_r;\mathbf{z}_r) = \log_2 \frac{| p\mathbf{H}_{r} \mathbf{H}_{r}^H + \mathbf{N}_r + \mathbf{Q}_r |}{| \mathbf{Q}_r |} 
\end{equation}
\begin{equation} \label{eqn:local_information_rate}
    I(\mathbf{x};\mathbf{z}_r) = \log_2 \frac{| p\mathbf{H}_{r} \mathbf{H}_{r}^H + \mathbf{N}_r + \mathbf{Q}_r |}{| \mathbf{N}_r + \mathbf{Q}_r |} 
\end{equation}
where we assume use of Gaussian signalling and a Gaussian quantization codebook. We can model lower complexity quantization schemes by adding a gap to the rate-distortion limit~\cite{Liu_Yu_Simeone_2017}; as we will show, this does not affect our approach to the problem. We refer to~\eqref{eqn:compression_rate} as the compression rate and~\eqref{eqn:local_information_rate} as the local information rate. In terms of joint processing at the CPU, we formulate the \textit{global information rate} as
\begin{equation} \label{eqn:global_information_rate}
    I(\mathbf{x};\mathbf{z}) = \log_2 \frac{| p\mathbf{H} \mathbf{H}^H + \mathbf{N} + \mathbf{Q} |}{| \mathbf{N} + \mathbf{Q} |}. 
\end{equation}

We wish to optimize this global information rate. To do so, we set the objective to maximize the theoretical network capacity expressed in terms of mutual information. Although achieving this information theoretical capacity requires a minimum mean square error successive interference cancellation (MMSE-SIC) receiver at CPU and joint process the signals at all RRHs in the network which is impractical and unscalable, it serves our goal of studying the theoretical performance of adaptive compression in fronthaul limited network.

As an aside, we note that robust formulations with imperfect CSI are well accepted, as in~\cite{li_2022, gamvrelis_2022}. Although we developed channel estimation techniques for our network model in~\cite{ammar_2022}, acquisition and sharing of CSI for distributed MIMO networks can take many forms~\cite{ammar_survey_2022} which should be carefully designed to decrease fronthaul overhead and maintain network performance. Thus, we will assume perfect CSI in our analysis and, in this paper, consider CSI sharing for the compression purposes only. The holistic architecture of channel estimation with limited fronthaul will be considered in future work.

\section{Optimal Local Compression} \label{section:optimal_local_compression}
Independently for each RRH, we define the local compression problem as finding the compression scheme, equivalently finding the optimal $\mathbf{Q}_r$, that maximizes the local information rate $I(\mathbf{x};\mathbf{z}_r)$ while satisfying the constraint that the compression rate does not exceed the fronthaul limiting capacity $I(\mathbf{y}_r;\mathbf{z}_r)\le L_r$. This is in contrast to the work in~\cite{wiffen_2021} which optimizes the transformation, but not the compression. We view $\mathbf{x} \rightarrow \mathbf{y}_r \rightarrow \mathbf{z}_r$ as a Markov chain where we call $\mathbf{x}$ the input, $\mathbf{y}_r$ the uncompressed output, and $\mathbf{z}_r$ the compressed output. Similar problems have been studied as Gaussian Information Bottleneck (GIB) problems in the field of machine learning and information theory~\cite{Chechik_2005,Winkelbauer_2014,wiffen_2019} and shown to have an optimal solution. Here, we arrive at the same result using a simpler and more intuitive analysis.

For any physical channel $\hat{\mathbf{H}}_r$, we can consider its corresponding eigen-channels $\mathbf{H}_r$ by applying the unitary EVD transform $\mathbf{U}_r$ such that $\mathbf{H}_r \mathbf{H}_r^H = \mathbf{U}_r^H\hat{\mathbf{H}}_r \hat{\mathbf{H}}_r\mathbf{U}_r$. We define $p\mathbf{H}_r\mathbf{H}_r^H=\mathbf{\Lambda}_r$ where $\mathbf{\Lambda}_r = \mathrm{diag} \{ \lambda_{r,m} \}_{m=1}^M$ is diagonal and the entries are in descending order. The data processing inequality ensures that this does not change the mutual information. We now define $r_{r,m}$ as the compression rate allocated to each eigen-channel, where $\sum_{m} r_{r,m} \le L_r$, and define $\mathbf{R}_r = \mathrm{diag} \{ 2^{-r_{r,m}} \}_{m=1}^M$. We use $2^{-r_{r,m}}$ because it is a finite, positive, and decaying function on $r_{r,m} \in [0,\infty)$.


We can now rewrite~\eqref{eqn:compression_rate} in scalar form in terms of compression rate of all eigen-channels as
\begin{align} 
& r_{r,m} = \log_2 \frac{ \lambda_{r,m} + \sigma_r^2 + q_{r,m} }{ q_{r,m} } \label{eqn:rate_in_terms_of_qerror_scalar} \\
\Rightarrow \hspace{0.05in} & q_{r,m} = \frac{ 2^{-r_{r,m}} (\lambda_{r,m} + \sigma_r^2) }{ 1 - 2^{-r_{r,m}}} .     \label{eqn:qerror_and_comprate_relation}
\end{align}
Using~\eqref{eqn:qerror_and_comprate_relation}, we can rewrite the mutual information in~\eqref{eqn:local_information_rate} as
\begin{equation} \label{eqn:local_information_rate_eigen}
    I(\mathbf{x};\mathbf{z}_r) = \log_2 \left| \mathbf{I}_M +  \frac{ \mathbf{\Lambda}_r ( \mathbf{I}_M - \mathbf{R}_r ) }{ \mathbf{N}_r + \mathbf{\Lambda}_r \mathbf{R}_r }  \right|
\end{equation}
where all matrices are diagonal and the division, used for notation convenience, is element-wise. There is an important intuition with the expression in~\eqref{eqn:local_information_rate_eigen}. The mutual information between the input and uncompressed output is $I(\mathbf{x};\mathbf{y}_r) = \log_2 |\mathbf{I} + \frac{\mathbf{\Lambda}_r}{\mathbf{N}_r} |$. We call $\frac{ \mathbf{N}_r (\mathbf{I}_M - \mathbf{R}_r) }{ \mathbf{N}_r + \mathbf{\Lambda}_r \mathbf{R}_r }$ as the penalty matrix, interpreted as multiplicative penalty from compression due to the limited fronthaul. With an unlimited fronthaul, $r_{r,m} \rightarrow \infty$ for each eigen-channel, i.e., $\mathbf{R}_r \rightarrow \mathbf{0}$ and the penalty matrix becomes an identity matrix. On the other hand, if the fronthaul limit approaches zero, we will have zero compression rate for each eigen-channel resulting in $\mathbf{R}_r \rightarrow \mathbf{I}_M$. The penalty matrix goes to zero resulting in zero local information rate. 

Since all matrices in~\eqref{eqn:local_information_rate_eigen} are diagonal, we can rewrite the local information rate function in scalar form as
\begin{equation} \label{eqn:local_information_rate_eigen_scalar}
    I(\mathbf{x};\mathbf{z}_r) = \sum_{m=1}^M \log_2 \frac{1 + \rho_{r,m}}{1 + \rho_{r,m} 2^{-r_{r,m}}}
\end{equation}
where $\rho_{r,m} = \lambda_{r,m} / \sigma_r^2$ is the signal-to-noise ratio (SNR) for each eigen-channel. Finally, we transform the local optimization problem that is $\mathbf{Q}_r$-based into a more intuitive compression rate allocation problem as
\begin{subequations}\label{eqn:local_compression_problem}
	\begin{align}
		\max_{ r_{r,1},\dots,r_{r,M} }\quad & I(\mathbf{x};\mathbf{z}_r)
		\label{subeqn:obj_func}
		\\
		\text{s.t.}\quad\quad
		& \sum_{m=1}^{M} r_{r,m} \le L_r
        \label{subeqn:capacity_constraint}
		\\
		& r_{r,m} \ge 0, \quad \forall m.
        \label{subeqn:non_negative_rate_constraint}
	\end{align}
\end{subequations}
for which a unique analytic solution exists. Using stationarity and complementary slackness of Karush-Kuhn-Tucker (KKT) conditions, we obtain the optimal compression rate allocation of each eigen-channel via waterfilling (WF) as
\begin{equation} 
    r_{r,m} = \left[ \log_2 \left( \frac{\rho_{r,m}(1-\nu)}{\nu}  \right) \right]^+ \label{eq:WF}
\end{equation}
with $[a]^+=\max(0,a)$ and $\nu$ being the Lagrange multiplier for constraint~\eqref{subeqn:capacity_constraint}. The water-level is $\log_2 \left( 1/\nu - 1 \right)$ and the ground-level is $\log_2 \left( 1/\rho_{r,m} \right)$ which can be reverse waterfilling if the signs are defined reversed. A more algorithmic convenient form can be written as 
\begin{equation} \label{eqn:waterfill_solution}
	r_{r,m} = \left[ \frac{L_r}{n_r} + \log_2(\rho_{r,m}) - \frac{1}{n_r} \sum_{m'=1}^{n_r} \log_2(\rho_{r,m'}) \right]^+
\end{equation}
with $n_r$ denoting the number of channels with non-zero rates. \eqref{eqn:waterfill_solution} is not only the optimal solution to~\eqref{eqn:local_compression_problem} but also to the original GIB problem optimized with respect to $\mathbf{Q}_r$ and can be rigorously shown by extending the results of~\cite{Chechik_2005}.

We remark that this local formulation does not necessarily maximize the global information rate $I(\mathbf{x};\mathbf{z})$. But it is simple to implement since it requires no cooperation between the RRHs and serves as a good performance indicator. Importantly, it provides the insight that the compression problem can be viewed as a dimension reduction problem that finds the optimal number of eigen-channels $n_r$ that have non-zero compression rate allocated to it. \eqref{eqn:waterfill_solution} represents a tradeoff: we want to decrease the dimension so fewer eigen-channels are sharing the limited fronthaul capacity; which results in more compression rates allocated to each channel so that they suffer less distortion. However, we also want to increase the output dimension to encourage multiplexing and interference cancellation among the users during processing at the CPU. The optimal dimension provides an optimal balance.

\section{Global Compression}
\subsection{Centralized Approach} \label{subsection:global_compression_centralized}
In the global picture, CPU finds the compression scheme that maximizes the global information rate~\eqref{eqn:global_information_rate}. A similar problem is addressed in~\cite{wiffen_2019}. Here, we will describe its procedure and add some intuition. Our goal is to use this result as a stepping stone for decentralized implementations.

Since we assume the channels at each RRH are uncorrelated into eigen-channels, the global objective can be rewritten as
\begin{equation} \label{eqn:global_information_rate_eigen}
	I(\mathbf{x};\mathbf{z}) = \log_2 \left| \mathbf{I}_{|\mathcal{U}|} + \sum_{r=1}^{|\mathcal{R}|} p \mathbf{H}_r^H \frac{  \mathbf{I}_M-\mathbf{R}_r }{ \mathbf{N}_r + \mathbf{\Lambda}_r \mathbf{R}_r  } \mathbf{H}_r  \right|
\end{equation}
which can be derived by substituting the quantization error covariance with $\mathbf{R}_r$ (similar to the local case) and using the matrix determinant lemma. The objective is to find the compression rate for all eigen-channels $r_{r,m}$ that maximizes~\eqref{eqn:global_information_rate_eigen} while satisfying the local fronthaul capacity constraint~\eqref{subeqn:capacity_constraint} and the non-negative rate constraint~\eqref{subeqn:non_negative_rate_constraint} for all RRHs.

Unfortunately, this cannot be formulated as a convex problem and we use projected gradient descent (PGD) to obtain a locally optimal solution. We emphasize that we do not claim PGD to be optimal; we select PGD to show validity of our formulations. We initialize with the local WF solution in~\eqref{eq:WF}, guaranteeing convergence to a solution that is better than the local method. Updates use the gradient given by
\begin{multline} \label{eqn:gradient_global_info_rate}
    \frac{\partial I(\mathbf{x};\mathbf{z})}{\partial r_{r,m}} = 
    \frac{ 2^{-r_{r,m}} (\sigma_r^2 + \lambda_{r,m} ) }{ ( \sigma_r^2 + \lambda_{r,m} 2^{-r_{r,m}} )^2 } \times
    \\ p \mathbf{h}_{r,m}^H \left( \mathbf{I}_{|\mathcal{U}|} + \sum_{r=1}^{|\mathcal{R}|} p \mathbf{H}_r^H \frac{  \mathbf{I}_M-\mathbf{R}_r }{ \mathbf{N}_r + \mathbf{\Lambda}_r \mathbf{R}_r  } \mathbf{H}_r \right)^{-1} \mathbf{h}_{r,m}
\end{multline}
where we denote $\mathbf{h}_{r,m}\in \mathbb{C}^{|\mathcal{U}|}$ as the column of $\mathbf{H}_r^T$. The projection is trivial because it is uncoupled between RRHs and also linear. We repeat the gradient and projection until reaches a local optimum. From simulations, the local optimum is heavily dependant on the initialization. Using WF as initialization is effective but cannot be guaranteed to be optimal.

We note that, compared to the local WF method, the global method tends to be more aggressive at reducing dimension, i.e., smaller $n_r$. One reason is that quantization distortion is seen globally than locally and another reason is that joint processing has more interference cancellation capabilities than the local approach and so a lower dimension is favored. 

\vspace{-0.1in}

\subsection{Decentralized Formulation} \label{subsection:decentralized_formulation}
With the global approach being impractical, our contribution is decentralized compression. Assuming the RRHs are equipped with RABs, we step toward this goal by rewriting the problem with local decisions on compression allocations. We first define $\mathbf{z}_{\backslash r}$ as the compressed signals for all $r'\in\mathcal{R}\backslash r$. The other variables are defined similarly where we use backslash to denote exclusion. The global objective can be written as $I(\mathbf{x};\mathbf{z}) = I(\mathbf{x};\mathbf{z}_{r}|\mathbf{z}_{\backslash r}) + I(\mathbf{x};\mathbf{z}_{\backslash r})$ with only the first term depending on $\mathbf{R}_r$. We note that an alternative to centralized compression is coordinate descent which, iteratively through all RRHs, maximizes $I(\mathbf{x};\mathbf{z}_{r}|\mathbf{z}_{\backslash r})$ with respect to $\mathbf{R}_r$ while keeping $\mathbf{R}_{\backslash r}$ fixed. 

The conditional mutual information is given by
\begin{equation} \label{eqn:conditional_mutual_information}
I(\mathbf{x};\mathbf{z}_{r}|\mathbf{z}_{\backslash r}) = \log_2 \left| \mathbf{I}_M+ p \mathbf{H}_r \mathbf{B}_{r}^{-1} \mathbf{H}_r^H  \frac{  \mathbf{I}_M-\mathbf{R}_r }{ \mathbf{N}_r + \mathbf{\Lambda}_r \mathbf{R}_r  }  \right|
\end{equation}
where $\mathbf{B}_{r}=\mathbf{I}_{|\mathcal{U}|}+ \sum_{r'\neq r} p \mathbf{H}_{r'}^H \frac{  \mathbf{I}_M-\mathbf{R}_{r'} }{ \mathbf{N}_{r'} + \mathbf{\Lambda}_{r'} \mathbf{R}_{r'}  } \mathbf{H}_{r'}$ is the side information matrix. The main challenge for decentralization is that maximizing~\eqref{eqn:conditional_mutual_information} cannot be done individually by one RRH as it is coupled with decisions on other RRHs.

One way to decouple the objectives is to approximate what is been done by the other RRHs~\cite{li_2024}. In our case, we need to approximate the side information matrix $\mathbf{B}_{r}$. A heuristic we use to approximate $\mathbf{R}_{\backslash r}$ is to assume they are doing local optimization with WF; we denote the resulting matrix as $\hat{\mathbf{B}}_r$. If we plug in $\hat{\mathbf{B}}_{r}$ in~\eqref{eqn:conditional_mutual_information}, we can already maximize the decentralized objective with PGD. However, we found the term $\mathbf{H}_r \hat{\mathbf{B}}_{r}^{-1} \mathbf{H}_r^H$ to be too inaccurate due to the approximations. To simplify this optimization further, we formulate the decentralized objective function as 
\begin{equation} \label{eqn:decentralized_objective}
I_{\Tilde{\mathbf{\Lambda}}_r}(\mathbf{R}_r) = \log_2 \left| \mathbf{I}_M+ \Tilde{\mathbf{\Lambda}}_r  \frac{  \mathbf{I}_M-\mathbf{R}_r }{ \mathbf{N}_r + \mathbf{\Lambda}_r \mathbf{R}_r  }  \right|
\end{equation}
where $\Tilde{\mathbf{\Lambda}}_r$ is the diagonal matrix from EVD of $p\mathbf{H}_r \hat{\mathbf{B}}_{r}^{-1} \mathbf{H}_r^H$. We can again use PGD. Even though~\eqref{eqn:decentralized_objective} is different from the approximation of~\eqref{eqn:conditional_mutual_information} with $\hat{\mathbf{B}}_{r}$,~\eqref{eqn:decentralized_objective} is less computationally heavy because its gradient only involves scalar operations where the gradient of~\eqref{eqn:conditional_mutual_information} includes matrix operations.

While we do not claim that this objective function maximizes the mutual information, our approach provides us a decentralized algorithm and also improves upon the local WF method by taking global information into account. We also refer to this as local compression with side information.

\vspace{-0.1in}

\subsection{Generalized Decentralized Compression}
We take a step back from the local case and study the case where we do not use a Gaussian quantizer. In such case, the compression rate can be written as~\cite{Liu_Yu_Simeone_2017}
\begin{equation} \label{eqn:compression_rate_non_gaussian_quantizer}
    I(\mathbf{y}_r;\mathbf{z}_r) = \log_2 \frac{| \Gamma_q \left( \mathbf{\Lambda}_r + \mathbf{N}_r \right) + \mathbf{Q}_r |}{| \mathbf{Q}_r |} 
\end{equation}
where $\Gamma_q$ denotes the gap to the rate-distortion limit. With procedures similar to Section~\ref{section:optimal_local_compression}, we can rewrite $\mathbf{Q}_r$ in terms of $\mathbf{R}_r$ using~\eqref{eqn:compression_rate_non_gaussian_quantizer} and substitute it in the rate in~\eqref{eqn:local_information_rate} obtaining
\begin{equation} 
    I(\mathbf{x};\mathbf{z}_r) = \log_2 \left| \mathbf{I}_M + \frac{ \mathbf{\Lambda}_r (\mathbf{I}_M - \mathbf{R}_r) }{ \mathbf{N}_r + ( \Gamma_q \mathbf{\Lambda}_r + (\Gamma_q - 1) \mathbf{N}_r ) ) \mathbf{R}_r }  \right| .
\end{equation}
Note that this equation is similar to~\eqref{eqn:local_information_rate_eigen} except for the denominator.  If we use a Gaussian quantizer, i.e., $\Gamma_q = 1$, they are the same. Connecting with how~\eqref{eqn:decentralized_objective} is also similar, this provides intuition to the two $\mathbf{\Lambda}_r$ in numerator and denominator of~\eqref{eqn:local_information_rate_eigen}. We formulate the generalized decentralized objective as
\begin{equation} \label{eqn:generalized_decentralized_objective}
I_{\mathbf{\Lambda}_r^{(i)},\mathbf{\Lambda}_r^{(c)}}(\mathbf{R}_r) = \log_2 \left| \mathbf{I}_M+   \frac{ \mathbf{\Lambda}_r^{(i)} ( \mathbf{I}_M-\mathbf{R}_r ) }{ \mathbf{N}_r + \mathbf{\Lambda}_r^{(c)} \mathbf{R}_r  }  \right|
\end{equation}
where $\mathbf{\Lambda}_r^{(i)}$ describes the amount of information depending on the processing schemes and $\mathbf{\Lambda}_r^{(c)}$ describes the penalization depending on the compression methods (can also be extended to joint compression such as Wyner-Ziv). Thus, the generalized decentralized optimization problem is 
\begin{subequations}\label{eqn:general_decentralized_compression_problem}
	\begin{align}
		\max_{ r_{r,1},\dots,r_{r,M} }\quad & I_{\mathbf{\Lambda}_r^{(i)},\mathbf{\Lambda}_r^{(c)}}(\mathbf{R}_r)
		\\
		\text{s.t.}\quad\quad
		& \sum_{m=1}^{M} r_{r,m} \le L_r, \hspace*{0.1in}
		r_{r,m} \ge 0, \quad \forall m
	\end{align}
\end{subequations}
and finds $\mathbf{\Lambda}_r^{(i)}$, $\mathbf{\Lambda}_r^{(c)}$ for different scenarios. For example, take $\mathbf{\Lambda}_r^{(c)} = \Gamma_q \mathbf{\Lambda}_r + (\Gamma_q - 1) \mathbf{N}_r$ for non-Gaussian quantizers and $\mathbf{\Lambda}_r^{(i)} = \Tilde{\mathbf{\Lambda}}_r$ when using Section~\ref{subsection:decentralized_formulation}.

\subsection{Scalable Computation of Side Information} \label{subsection:scalable_method}
In Section~\ref{subsection:decentralized_formulation}, we formulated the decentralized objective by computing an approximation of the side information matrix $\mathbf{B}_{r}$. Recall that this matrix is given by
\begin{equation} \label{eqn:side_information_matrix}
    \mathbf{B}_{r}=\mathbf{I}_{|\mathcal{U}|}+ \sum_{r' \in \mathcal{R}\backslash r} p \mathbf{H}_{r'}^H \frac{  \mathbf{I}_M-\mathbf{R}_{r'} }{ \mathbf{N}_{r'} + \mathbf{\Lambda}_{r'} \mathbf{R}_{r'}  } \mathbf{H}_{r'}
\end{equation}
which assumes RRH $r$ knows the channels $\mathbf{H}_{r'}$ for RRH $r'$. An efficient way to implement this is to have the CPU compute these matrices and send them to designated RRHs. However, this would cost an additional overhead of $|\mathcal{U}|^2$ complex numbers on each fronthaul link for every coherence time since the matrices need to be recomputed if channels change. In essence, our goal is to find compact representations of $\mathbf{B}_{r}$ that is used to find $\mathbf{\Lambda}_r^{(i)}$; we will focus on two methods.

The first, ``Statistical CSI'', method approximates the channels with large scale statistics, including pathloss and shadowing. From Section~\ref{subsection:network_model}, we define $\mathbb{E}\{ p \mathbf{h}_{ru} \mathbf{h}_{ru}^H \} = \mathbf{\Psi}_{ru}^{(s)} $ and $\mathbb{E}\{ p \mathbf{H}_{r} \mathbf{H}_{r}^H \} = \mathbf{\Psi}_{r}^{(s)} $ where $\mathbf{\Psi}_{r}^{(s)} = \sum_{u \in \mathcal{U}}\mathbf{\Psi}_{ru}^{(s)}$. 

Computing~\eqref{eqn:side_information_matrix} also requires $\mathbf{\Lambda}_{r'}$ and $\mathbf{R}_{r'}$ which implicitly requires CSI. To avoid this, we \textit{approximate} non-local compression strategies, denoted as $\hat{\mathbf{R}}_{r'}^{(s)}$, as equally splitting the compression rate across all dimensions. We can estimate the penalty matrix as $\mathbf{P}_{r'}^{(s)} = \frac{  \mathbf{I}_M-\hat{\mathbf{R}}_{r'}^{(s)} }{ \mathbf{N}_{r'} + \mathbf{\Psi}_{r'}^{(s)} \hat{\mathbf{R}}_{r'}^{(s)}  }$. Finally, we arrive at $\mathbb{E}\{ p \mathbf{H}_{r'}^H \mathbf{P}_{r'}^{(s)} \mathbf{H}_{r'} \} = \mathrm{diag} \{ \mathrm{tr}( \mathbf{\Psi}_{r'u}^{(s)} \mathbf{P}_{r'}^{(s)} ) \}_{u=1}^{|\mathcal{U}|} $ which we can used	 to compute
\begin{equation} \label{eqn:side_information_matrix_statistical}
    \hat{\mathbf{B}}_{r}^{(s)}=\mathbf{I}_{|\mathcal{U}|}+ \sum_{r' \in \mathcal{R}\backslash r} \mathrm{diag} \{ \mathrm{tr}( \mathbf{\Psi}_{r'u}^{(s)} \mathbf{P}_{r'}^{(s)} ) \}_{u=1}^{|\mathcal{U}|}
\end{equation}
on the CPU then shared with corresponding RRHs. 

There are a few aspects that makes this procedure scalable. Though~\eqref{eqn:side_information_matrix_statistical} is written in matrix form, all components are scaled identity matrices, so only scalar operations are involved. In addition, it only costs an additional overhead of $|\mathcal{U}|$ on each fronthaul link. Most importantly, the procedure only depends on large scale effects which change slowing, meaning less frequent computation and reduced fronthaul overhead overall.

The second, ``Traffic Distribution'', method we consider only requires knowing the user traffic density. We define $\Upsilon (x,y)$ as the traffic probability density function (PDF) which can be constructed from a traffic survey within the region and $x$, $y$ being the 2D coordinates. For example, this PDF can be modelled as a mix of a uniform distribution and some hotspots modeled as bivariate normal distributions in~\cite{li_2022,ammar_2022_distributed}. We can approximate the contribution of each user on each RRH as
\begin{multline} \label{eqn:traffic_integration}
    [ \mathbf{\Psi}_{ru}^{(t)} ]_{mm}  = \iint_{x,y} p \Upsilon (x,y) \\ \beta\left( d_{\mathrm{excl}}+\sqrt{(x-x_r)^2+(y-y_r)^2} \right)  \mathrm{d}x\mathrm{d}y
\end{multline}
with $x_r$, $y_r$ denote the coordinate of RRH $r$, $d_{\mathrm{excl}}$ denotes a small exclusion distance, and the integration is done over the whole network. Since~\eqref{eqn:traffic_integration} is independent of $m$ and $u$, we can treat $\mathbf{\Psi}_{ru}^{(t)}$ as a scalar times identity and $\mathbf{\Psi}_{r}^{(t)} = |\mathcal{U}| \mathbf{\Psi}_{ru}^{(t)}$.

Using the same procedure as before, we use equal compression rate approximations $\hat{\mathbf{R}}_{r'}^{(t)}$, obtain penalty matrix $\mathbf{P}_{r'}^{(t)}$, and finally arrive at $\hat{\mathbf{B}}_{r}^{(t)}$ with the same formulation in~\eqref{eqn:side_information_matrix_statistical} but different superscripts. The key difference is that, because we have not acquired any channel statistics and all users share the same traffic PDF, $\hat{\mathbf{B}}_{r}^{(t)}$ is proportional to an identity matrix. As a result, the CPU only needs to send one scalar to the corresponding RRH over the fronthaul. Most importantly, the user density changes even less frequently than large scale statistics, suggesting that the overhead may become negligible.

Despite the many approximations done for the two methods, the performance of the decentralized compression rate allocation is barely affected. One reason is that the main penalty comes from estimating $\mathbf{R}_{\backslash r}$ and treating it as static due to the need of decentralization but not from approximating the channels. This is because, as discussed before, the global method is more aggressive at reducing the dimensions than local WF. For small $n_r$, the water-level is higher, making perturbations to $r_{r,m}$ having less impact; so, the problem becomes dimension $n_r$ dominant. Decentralization essentially becomes applying the right scaling to $\mathbf{\Lambda}_r$ so that the optimizer for~\eqref{eqn:generalized_decentralized_objective} lands in the best dimension. This is easier than finding the exact rate. Hence, we have some wiggle room for approximations making scalability possible.

\section{Numerical Results and Discussion}
In this section, we present the results of simulations to illustrate the efficacy of the proposed compression strategies. We consider a wrap-around structure with 7 hexagonal cells. RRHs are uniformly distributed in each cell and are connected to one CPU. Each cell has a radius of 400 m and users are uniformly distributed with a 20 m exclusion region around the RRHs. We use the COST231 Walfisch-Ikegami model~\cite{Walfisch_1988} to define the path loss component at the $f = 1800$ MHz band as $\beta(d_{ru}) = -112.4271 - 38 \mathrm{log}_{10} (d_{ru})$, where $d_{ru}$ is measured in km, and a 4 dB lognormal shadowing.

\begin{figure}
	\centering
	\includegraphics[width=0.9\linewidth]{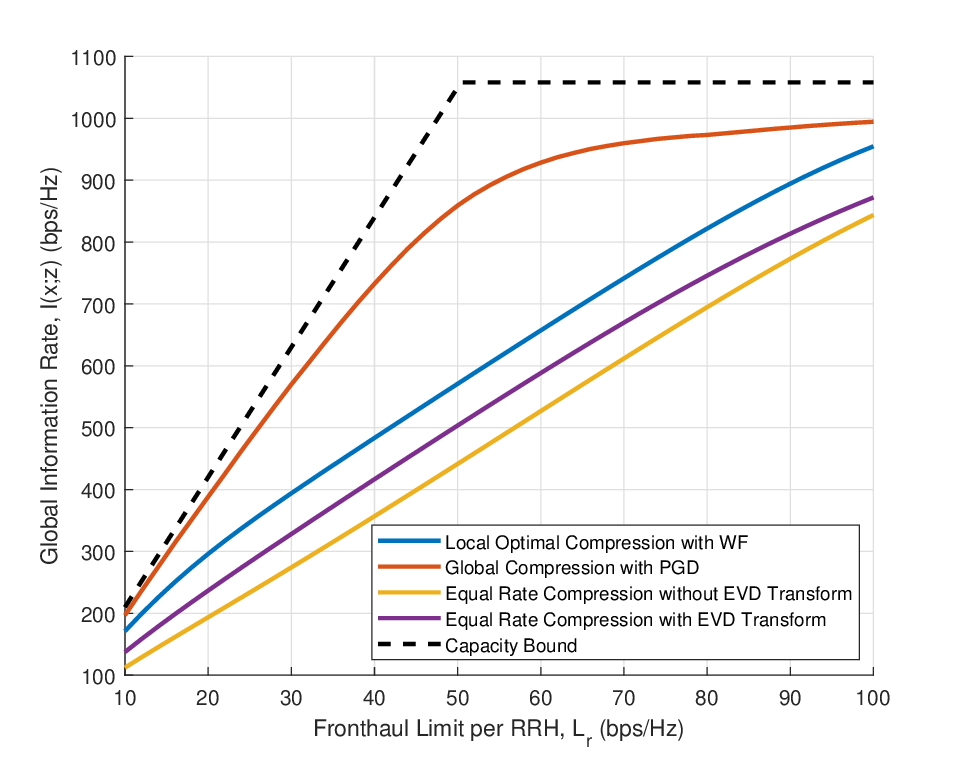}
	\caption{Comparison between local and global methods with global information rate $I(\mathbf{x};\mathbf{z})$ as a function of fronthaul limit per RRH $L_r$.}
	\label{fig:local_vs_global}
\end{figure}

For our simulation, we consider 10 single-antenna users and 3 RRHs per cell with each RRH equipped with 8 antennas. We run a Monte-Carlo simulation and average our results over 100 topologies. We analyze the performance of different compression methods based on the global information rate or capacity, $I(\mathbf{x};\mathbf{z})$, as a function of fronthaul limit per RRH, $L_r$. We also include a cut-set bound for capacity to show the best that can be achieved without a fronthaul limit. For example, in Fig.~\ref{fig:local_vs_global}, the slanted line represents the sum of fronthaul limit for all RRHs whereas the horizontal line represents the information rate without compression $I(\mathbf{x};\mathbf{y})$. We assume a Gaussian quantizer with $\Gamma_q=1$. Though we set all the fronthaul links to have the same limit, our methods also work for the case of different limits for each link. This is because our rate-based method finds a RRH-specific dimension reduction strategy for each RRH, which automatically considers the case of different $L_r$. Other transform-based methods~\cite{Liu_Yu_Simeone_2017,wiffen_2021} that fix the same arbitrary dimension to reduce to for all RRHs will be penalized by heterogeneous fronthauls. 

In Fig.~\ref{fig:local_vs_global}, we simulated 4 different compression methods. The first and second methods (blue and red) are local optimal compression with WF and global compression with PGD where the procedures are described in Sections~\ref{section:optimal_local_compression} and~\ref{subsection:global_compression_centralized}. The third method (yellow) does not apply the EVD transform and the compression rates $r_{r,m}$ are split equally among the physical channels. The fourth method (purple) applies EVD and the compression rates are evenly split among the eigen-channels. Despite equal compression on physical channel is the most practical strategy, it requires a significant amount of fronthaul to reach the cut-set bound which is the main limitation to compress-forward schemes. There is an increase from the yellow to purple curves suggesting the EVD improves compression by exploiting correlations of the received signals. The increase from the purple to blue curves suggests efficacy of dimension reduction. Though the improvement seems insignificant, it is optimal if we are evaluating based on $\sum_{r\in\mathcal{R}} I(\mathbf{x};\mathbf{z}_r)$. Finally, there is a significant increase from blue to red which shows the benefit of considering global knowledge which motivates us to find decentralized and practical versions of global compression.

\begin{figure}
	\centering
	\includegraphics[width=0.9\linewidth]{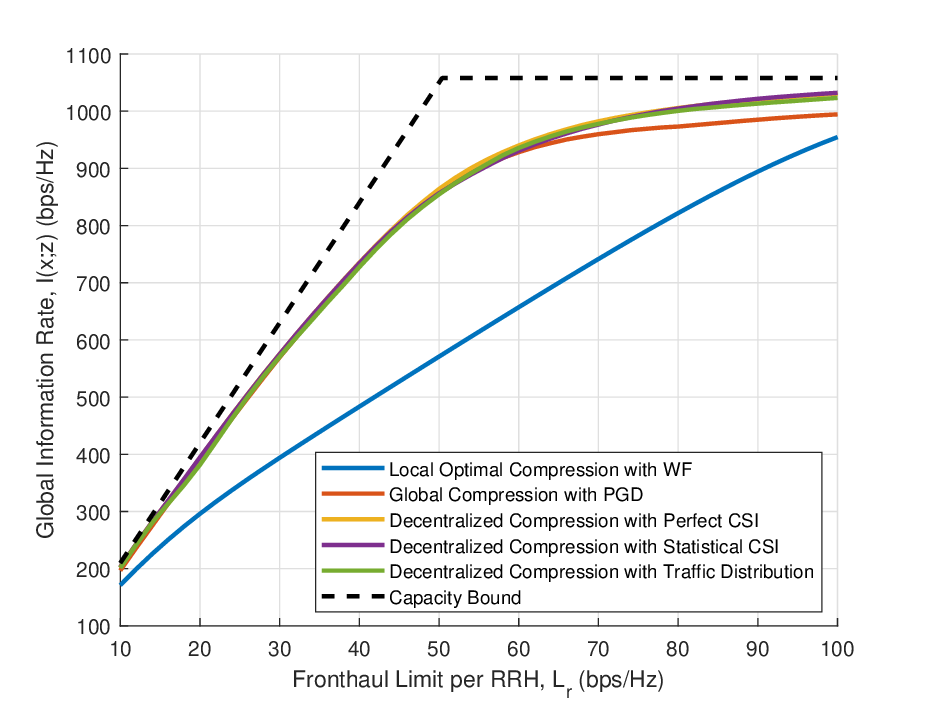}
	\caption{Comparison between decentralized methods with global information rate $I(\mathbf{x};\mathbf{z})$ as a function of fronthaul limit per RRH $L_r$.}
	\label{fig:decentralized_comparison}
\end{figure}

In Fig.~\ref{fig:decentralized_comparison}, we use the local WF method and global PGD methods as reference and compare the performance of the three decentralized methods proposed in Sections~\ref{subsection:decentralized_formulation} and~\ref{subsection:scalable_method}. Surprisingly, the decentralized methods that use the different approximations outperforms centralized PGD at high $L_r$. This is because we used local WF for initialization and PGD is often stuck in local optima. We tried different initializations that perform better, but they are ad hoc and local WF is the best without prior knowledge. Thus, the decentralized methods help us jump out of local optima, partly because the decentralized objective in~\eqref{eqn:decentralized_objective} has fewer degrees of freedom than the centralized objective in~\eqref{eqn:global_information_rate_eigen}, making it easier to deal with. Finally, and most importantly, the approximations do not effect the performance of decentralized methods on average. This suggests, with RABs at RRHs and using the Traffic Distribution method, we can find good compression strategies that not only utilize global information but also incur minimal fronthaul overhead.

\section{Conclusions}\label{section:conclusion}
This paper studied the theoretical achievable information rate of adaptive compression problem for the uplink cell-free MIMO network under limited fronthaul. We used WF with respect to the eigen-channels to find the optimal compression rate allocation when only local CSI is available and we reached upon an insight that the rate-based approach indirectly suggests the dimension reduction strategy. Then extending on global mutual information, we formulated the decentralized compression problem that can be solved locally at each RRH with PGD. Though decoupled, the RABs still require global CSI as side information to be shared which cause significant overhead. As a result, we proposed two methods, Statistical CSI and Traffic Distribution, reducing the overhead significantly by representing side information more compactly and also decreasing the frequency of sharing.

Our numerical results show that utilizing global CSI will outperform purely local methods, approaching the cut-set bound. Surprisingly, the decentralized method outperforms the centralized method since the latter suffers from local optima. Importantly, the Traffic Distribution method achieves the same network capacity as others, with minimal communication overhead. This suggests it is a promising solution for scalable decentralized adaptive compression.


\bibliography{biblio}
\bibliographystyle{ieeetr}

\end{document}